\documentclass[prl,reprint,amsmath,amssymb,aps,preprintnumbers,showpacs,twocolumn]{revtex4}
\usepackage[colorlinks=true,linkcolor=black, citecolor=black,urlcolor=black]{hyperref} 

\begin{document}

\vspace*{0cm}

\title{Condition on Ramond-Ramond fluxes for factorization \\of worldsheet scattering in anti-de Sitter space}
\author{Linus Wulff}
\email{wulff@physics.muni.cz}

\affiliation{Department of Theoretical Physics and Astrophysics, Masaryk University, 611 37 Brno, Czech Republic}

\begin{abstract}
Factorization of scattering is the hallmark of integrable 1+1 dimensional quantum field theories. For factorization of scattering to be possible the set of masses and momenta must be conserved in any two-to-two scattering process. We use this fact to constrain the form of the Ramond-Ramond fluxes for integrable supergravity anti-de Sitter backgrounds by analysing tree-level scattering of two AdS bosons into two fermions on the worldsheet of a BMN string. We find a condition which can be efficiently used to rule out integrability of AdS strings and therefore of the corresponding AdS/CFT dualities, as we demonstrate for some simple examples.
\end{abstract}


\maketitle

\section{Introduction}

A key to understanding and checking precisely the AdS/CFT correspondence~\cite{Maldacena:1997re}, which relates string theory in $(d+1)$-dimensional anti-de Sitter (AdS) backgrounds to conformal field theories in $d$ dimensions, has been the discovery of integrability on both sides of the correspondence. Originally for the superstring on $AdS_5\times S^5$~\cite{Bena:2003wd} and its dual $\mathcal{N}=4$ super Yang-Mills theory in four dimensions~\cite{Minahan:2002ve}, see the reviews~\cite{Beisert:2010jr,Bombardelli:2016rwb}. In particular this has allowed for computing the spectrum of the quantum theory in the large $N$ limit \emph{exactly}~\cite{Bombardelli:2009ns,Gromov:2009bc,Arutyunov:2009ur,Gromov:2013pga}.

Other AdS/CFT examples have been found which also possess an integrable structure \cite{Aharony:2008ug,Arutyunov:2008if,Stefanski:2008ik,Sorokin:2010wn,Babichenko:2009dk,Sorokin:2011rr,Wulff:2014kja}. An interesting question is whether there are more such examples out there to be found. In order to begin to tackle this question we will derive constraints on the supergravity AdS background where the string propagates which are needed for integrability. There are many ways to do this, e.g. \cite{Basu:2011di,Stepanchuk:2012xi,Chervonyi:2013eja,Giataganas:2013dha}. Here we will follow an approach similar to \cite{Wulff:2017hzy}. The idea is to expand around a classical string solution where one has a notion of a worldsheet S-matrix. This S-matrix is required to be of factorized form, i.e. to reduce to a sequence of two-to-two scattering events, in the case of an integrable theory and this places very strong constraints on its form \cite{Zamolodchikov:1978xm}. In particular factorization of the S-matrix requires the set of masses and momenta to be conserved in any two-to-two scattering process. Here we will expand around the BMN string and compute scattering of two worldsheet bosons, coming from the transverse fluctuations in AdS, into two worldsheet fermions. The reason we want to include fermions is that in many examples we are interested in, e.g. symmetric spaces, the bosonic string sigma model is integrable. We expect that integrability is lost once fermions are included unless the background is supersymmetric \cite{Wulff:2017hzy}. The mass of the AdS bosons is 1 in suitable units while the mass spectrum of the fermions is determined by the fluxes. Factorization then implies that unless the fermions also have mass 1 this pair production amplitude must vanish.

As we will see this puts strong constraints on the fluxes of the background (we will simplify things by assuming the NSNS flux does not contribute, when it does it is already strongly constrained at the bosonic level). The constraint we find on the RR fluxes is given in (\ref{eq:int-cond}) with the RR fluxes encoded in two constant matrices $M$ and $N$ via (\ref{eq:MN}) and (\ref{eq:SIIA}) or (\ref{eq:MN-IIB}) and (\ref{eq:SIIB}). The matrix $M$ determines the fermion mass spectrum while $N$ determines the relevant Yukawa couplings. This condition does not depend on any of the particle momenta and arises by taking a limit of large centre-of-mass energy.

We first recall the superstring action and its near-BMN expansion and gauge fixing. Then we compute the amplitude for scattering of two identical AdS bosons into two fermions and derive the constraint on the RR fluxes. Finally we apply this constraint to rule out integrability for some symmetric space backgrounds.

\section{String action and near-BMN expansion}
Our starting point is the Green-Schwarz superstring Lagrangian \cite{Cvetic:1999zs,Wulff:2013kga}\footnote{Here $\gamma^{ij}=\sqrt{-h}h^{ij}$ with $h_{ij}$ the worldsheet metric with signature $(-,+)$, $\varepsilon^{01}=1$ and $\theta\Gamma^a\mathcal D\theta=\theta^\alpha\mathcal C_{\alpha\beta}(\Gamma^a)^\beta{}_\gamma(\mathcal D\theta)^\gamma$. When working in conformal gauge one also has to include the Fradkin-Tseytlin counter-term $\Phi R^{(2)}$ \cite{Fradkin:1984pq,Fradkin:1985ys} where $\Phi$ is the dilaton superfield, whose expansion to quadratic order in $\theta$ can be found in \cite{Wulff:2013kga}, see \cite{Wulff:2016tju}. Here we will work at tree-level and in light-cone gauge so this term will not be relevant.}
\begin{align}
\mathcal L=&
-\tfrac{T}{2}e_i{}^ae_j{}^b(\gamma^{ij}\eta_{ab}-\varepsilon^{ij}B_{ab})
\nonumber\\
&{}
-iTe_i{}^a\,\theta\Gamma_a(\gamma^{ij}-\varepsilon^{ij}\Gamma_{11})\mathcal D_j\theta
+\mathcal O(\theta^4)\,,
\label{eq:S}
\end{align}
where the derivative operator is the same that appears in the Killing spinor equation, namely
\begin{equation}
\mathcal D=d-\tfrac14\omega^{ab}\Gamma_{ab}+\tfrac18e^a(H_{abc}\Gamma^{bc}\Gamma_{11}+\mathcal S\Gamma_a)\,.
\label{eq:D}
\end{equation}
The $\theta^4$-terms are also know \cite{Wulff:2013kga} but they won't be needed here. The bosonic fields appearing here are the pull-backs to the worldsheet of type II supergravity fields -- the vielbeins $e^a$ ($a=0,\ldots,9$) and spin connection $\omega^{ab}$, the NSNS two-form $B_{ab}$ and its field strength $H=dB$ and the RR field strengths encoded in the bispinor $\mathcal S$. We follow the conventions of \cite{Wulff:2013kga} and the action we have written is for the type IIA superstring with $\theta$ a 32-component Majorana spinor and
\begin{equation}
\mathcal S=e^\phi(F^{(0)}+\tfrac12F^{(2)}_{ab}\Gamma^{ab}\Gamma_{11}+\tfrac{1}{4!}F^{(4)}_{abcd}\Gamma^{abcd})\,,
\label{eq:SIIA}
\end{equation}
in terms of the dilaton $\phi$ and RR field strengths. The action for the type IIB superstring is obtained by the replacements $\Gamma^a\rightarrow\gamma^a$, $\Gamma_{11}\rightarrow\sigma^3$ and
\begin{equation}
\mathcal S=-e^\phi(F^{(1)}_ai\sigma^2\gamma^a+\tfrac{1}{3!}F^{(3)}_{abc}\sigma^1\gamma^{abc}+\tfrac{1}{2\cdot5!}F^{(5)}_{abcde}i\sigma^2\gamma^{abcde})\,,
\label{eq:SIIB}
\end{equation}
where $\theta$ now consists of two 16-component Majorana-Weyl spinors of the same chirality and $\gamma^a$ are 16-component gamma matrices while the Pauli matrices mix the two spinors. For more details see the appendix of \cite{Wulff:2013kga}.

We are interested in backgrounds of the form $AdS_n\times M_{10-n}$ and we take the AdS metric to have a convenient form for light-cone gauge fixing
\begin{equation}
ds^2_{AdS}=R^2\left(-\left(\frac{1+\tfrac14z_m^2}{1-\tfrac14z_m^2}\right)^2dt^2+\frac{dz_m^2}{(1-\tfrac14z_n^2)^2}\right)\,,
\end{equation}
with spin connection
\begin{equation}
\omega^{0m}=-\tfrac12z^m(R^{-1}e^0+dt)\,,\qquad\omega^{mn}=R^{-1}z^{[m}e^{n]}\,,
%
\end{equation}
where $R$ is the AdS radius and $z_m$ ($m=1,\ldots,n-1$) are the transverse AdS coordinates. We assume that $M_{10-n}$ has a $U(1)$ isometry (it does not need to be compact) generated by a geodesic so that the metric can be written
\begin{equation}
ds^2_M=G_{m'n'}dy^{m'}dy^{n'}+G_{m'}dy^{m'}dx^9+Gdx^9dx^9\,,
%
%
%
%
\end{equation}
where $y^{m'}$ ($m'=n,\ldots,8$) are the transverse coordinates of $M_{10-n}$ and $G_{m'n'}$, $G_{m'}$ and $G$ are functions of these satisfying $G(0)=R^2$ and $G_{m'}(0)=\partial_{m'}G(0)=0$ while $x^9$ is the coordinate of the $U(1)$ isometry (suitably normalized). The condition that the linear term in $G(y)$ be absent is needed for the isometry to be a geodesic. The other two conditions can be arranged by rescaling and shifting $x^9$.

We also assume that the NSNS three-form $H$ has no legs in the 0,1 or 9-directions and that $\mathcal S$, encoding the RR fluxes, is independent of $t,z_1$ and $x^9$ and respects the ''boost invariance'' in the 1-direction, i.e. $[\Gamma^{01},\mathcal S]=0$. The assumptions involving the 1-direction are not necessary but they will simplify the analysis.

These conditions guarantee that there exist a BMN solution \cite{Berenstein:2002jq} of the string equations of motion taking the form
\begin{equation}
x^+=\tfrac12(x^0+x^9)=\tau\,,
\end{equation}
with $\tau$ the worldsheet time-coordinate. We expand the string Lagrangian (\ref{eq:S}) around this solution fixing so-called uniform light-cone gauge
\begin{equation}
x^+=\tau\,,\qquad
\frac{\partial\mathcal L}{\partial \dot x^-}=-2g\,,\qquad
\frac{\partial\mathcal L}{\partial x'^-}=0\,,
\label{eq:gauge-b}
\end{equation}
where we have defined the dimensionless coupling $g=TR^2$. The last two conditions on the momentum density conjugate to $x^-$ remove the two degrees of freedom of $\gamma^{ij}$. The Virasoro constraints remove the degrees of freedom associated to $x^-$. The kappa gauge invariance of the fermions is fixed by the corresponding condition
\begin{equation}
\Gamma^+\theta=0\quad\Leftrightarrow\quad \theta=P_+\theta\,,\quad P_\pm=\tfrac12(1\pm\Gamma^{09})\,,
\label{eq:gauge-f}
\end{equation}
where $\Gamma^\pm=\tfrac12(\Gamma^0\pm\Gamma^9)$.

Since we will be interested here only in tree-level $z_1z_1\rightarrow\theta\theta$ scattering we will only keep the terms which can contribute to this. Our assumption that $H$ has no legs in the $0,1$ or 9-directions implies that (up to total derivatives) there cannot be any cubic couplings of the form $yz_1z_1$ coming from the $B$-field. Therefore the only contributions to $z_1z_1\rightarrow\theta\theta$ scattering at tree-level come from terms of the form $z_1\theta\theta$ and $z_1z_1\theta\theta$. Setting all the bosons except $z_1$ to zero the gauge fixing conditions in (\ref{eq:gauge-b}) lead to $\gamma^{ij}=\eta^{ij}+\hat\gamma^{ij}$ with \footnote{Note that the kappa gauge-fixing (\ref{eq:gauge-f}) implies that there are no $dx^-\theta\theta$-terms only $dx^-z_1^2\theta\theta$-terms, which cannot contribute at the order we are interested in.}
\begin{equation}
\hat\gamma^{00}=\hat\gamma^{11}=\tfrac12z_1^2+\ldots\,,\qquad\hat\gamma^{01}=0+\ldots\,,
\end{equation}
where the ellipsis denotes terms which cannot contribute to the order we are interested in. Using this in (\ref{eq:S}) and noting that the conditions on $H$ guarantee that it does not contribute while the spin connection also drops out one finds the Lagrangian
\begin{align}
\mathcal L
=&{}
\tfrac12\partial_+z_1\partial_-z_1
-\tfrac12z_1^2
-\tfrac{i}{2}\theta_+\Gamma^-\partial_+\theta_+
-\tfrac{i}{2}\theta_-\Gamma^-\partial_-\theta_-
\nonumber\\
&{}
-\theta_+\Gamma^{01}M\theta_-
-\tfrac{i}{2\sqrt g}(\partial_+z_1-\partial_-z_1)\,\theta_+\Gamma^0N\Gamma^1\theta_-
\nonumber\\
&{}
+\tfrac{i}{8g}z_1^2\big(\theta_+\Gamma^-\partial_-\theta_++\theta_-\Gamma^-\partial_+\theta_-\big)
\nonumber\\
&{}
+\tfrac{1}{4g}\partial_+z_1\partial_-z_1\,\theta_+\Gamma^{01}M\theta_-
+\ldots
\label{eq:L}
\end{align}
Here we have rescaled the fields as $z_1\rightarrow g^{-1/2}z_1$, $\theta\rightarrow\frac12R^{1/2}g^{-1/2}\theta$. We have also defined $\partial_\pm=\partial_0\pm\partial_1$ and $\theta_\pm=\frac12(1\pm\Gamma_{11})\theta$ and used our assumption that $\Gamma^{01}$ commutes with $\mathcal S$ to simplify the cubic terms. Furthermore we have split $\mathcal S$ into matrices $M$ and $N$ which commute with $\Gamma^0$, $\Gamma^9$ and $\Gamma_{11}$ defined by
\begin{equation}
P_+\mathcal SP_-|=\tfrac{4i}{R}\Gamma^{01}MP_-\,,\quad P_+\mathcal SP_+|=\tfrac{4}{R}NP_+\,.
\label{eq:MN}
\end{equation}
The vertical bar means that $\mathcal S$ is evaluated setting $z_m=y_{m'}=0$ so that $M$ and $N$ are constant matrices. It follows from the anti-symmetry of $\mathcal S$ that they satisfy $M^T=\Gamma^1M\Gamma^1$ and $N^T=-\Gamma^1N\Gamma^1$. We have written things so that the type IIB case is obtained by replacing $\Gamma^a\rightarrow\gamma^a$ \emph{and} $M\rightarrow iM$ and $N\rightarrow iN$ in (\ref{eq:L}) where now $M$ and $N$ are defined as
\begin{equation}
P_+\mathcal SP_-|=\tfrac{4}{R}\gamma^{01}MP_-\,,\quad P_+\mathcal SP_+|=\tfrac{4i}{R}NP_+
\label{eq:MN-IIB}
\end{equation}
and \emph{anti}-commute with $\gamma^0$, $\gamma^9$ and $\sigma^3$. They satisfy $M^T=-\gamma^1M\gamma^1$ and $N^T=\gamma^1N\gamma^1$.

Looking at the Lagrangian (\ref{eq:L}) we see that the AdS boson $z_1$ has mass 1 while the fermion mass spectrum is determined by the matrix $M$. The matrix $N$ encodes the Yukawa-type couplings.
We will now consider $z_1z_1\rightarrow\theta\theta$ scattering. Unless the fermions also have mass 1 this amplitude must vanish in an integrable theory to be compatible with factorized scattering.

%

\section{Tree-level $\mathbf{zz\rightarrow\theta\theta}$ scattering}
We find it convenient to work directly with the 8-component spinors $\theta_\pm$. The propagator takes the form
\begin{equation}
\langle\theta_\pm\theta_\pm\rangle
=
\left(
\begin{array}{cc}
 k_- & -\Gamma^1M\\
-\Gamma^1M & k_+
\end{array}
\right)\frac{i\Gamma^+}{k_+k_--M^TM}
\,.
\end{equation}
External state fermions come with factors of
\begin{equation}
u^i_\pm(k)=\left(
\begin{array}{c}
\sqrt{k_-}u^i\\
-m_i^{-1}\sqrt{k_+}\Gamma^1Mu^i
\end{array}
\right)\,,
\end{equation}
solving the free Dirac equation. Here $u^i$, with $i,j=1,\ldots,8$ labelling the eight physical fermions, is a constant (commuting) spinor which we take to be a suitably normalized eigenstate of the mass-squared operator 
\begin{equation}
M^TMu^i=m_i^2u^i\,,\quad
u^i\Gamma^-u^j=\delta^{ij}\,.
\end{equation}
In the case of type IIB we have $i\gamma^1$ in place of $\Gamma^1$ in the above expressions. Throughout the remainder of this section the type IIB expressions are obtained simply by setting $\Gamma^-\rightarrow\gamma^-$.

We are now ready to compute the amplitude for $z_1z_1\rightarrow\theta\theta$ scattering. The contribution from the quartic interaction terms in (\ref{eq:L}) is the simplest. It takes the form
\begin{align}
\mathcal A_4^{ij}=&
\tfrac{i}{4g}\delta^{ij}
\Big[
(p_{3+}-p_{4+})\sqrt{p_{3+}p_{4+}}
-(p_{3-}-p_{4-})\sqrt{p_{3-}p_{4-}}
\nonumber\\
&{}
+m_i(p_{1+}p_{2-}+p_{1-}p_{2+})\big(\sqrt{p_{3-}p_{4+}}-\sqrt{p_{3+}p_{4-}}\big)
\Big]\,.
\end{align}
Using the on-shell conditions $p_{1-}=1/p_{1+}$, $p_{2-}=1/p_{2+}$, $p_{3-}=m_i^2/p_{3+}$, $p_{4-}=m_j^2/p_{4+}$ and energy-momentum conservation, which implies for example (for $m_i=m_j$) that $m_i^2p_{1+}p_{2+}=p_{3+}p_{4+}$, this becomes
\begin{equation}
\mathcal A_4^{ij}=\tfrac{im_i}{4g}\delta^{ij}(p_{3+}-p_{4+})(1-p_{1+}^2)(1-p_{2+}^2)(p_{1+}p_{2+})^{-3/2}\,.
\label{eq:A4-final}
\end{equation}
We could of course express the amplitude in terms of only the incoming momenta $p_1$ and $p_2$ \footnote{
We have
$
p_{3+}-p_{4+}
=
(p_{1+}+p_{2+})
\times
\left[
(m_i^2-m_j^2)s^{-1}
+\sqrt{[(m_i^2-m_j^2)s^{-1}+1]^2-4m_i^2s^{-1}}
\right]
$ where $s=-(p_1+p_2)^2=(p_{1+}+p_{2+})^2/(p_{1+}p_{2+})$.
} but we have kept the factor of $(p_{3+}-p_{4+})$ to avoid complicating the expression too much.

Now we turn to the contribution from the cubic interaction terms in (\ref{eq:L}). This contribution is somewhat more complicated and takes the form
\begin{align}
\mathcal A_3^{ij}=&\tfrac{i}{4g}
(p_{1+}-p_{1-})(p_{2+}-p_{2-})
\nonumber\\
&\times\Big[u^i\Gamma^-\mathcal M'u^j-u^j\Gamma^-\mathcal M'(p_3\leftrightarrow p_4)u^i\Big]\,,
\end{align}
where
\begin{align}
\mathcal M'=&
NM\frac{m_j^{-1}\sqrt{p_{3-}p_{4+}}}{-(p_1-p_3)^2-M^TM}NM
\\
&{}
-M^TN^T\frac{m_i^{-1}\sqrt{p_{3+}p_{4-}}}{-(p_1-p_3)^2-M^TM}M^TN^T
\nonumber\\
&{}
+N\frac{(p_{1+}-p_{3+})\sqrt{p_{3-}p_{4-}}}{-(p_1-p_3)^2-MM^T}N^T
\nonumber\\
&{}
-M^TN^T\frac{(m_im_j)^{-1}(p_{1-}-p_{3-})\sqrt{p_{3+}p_{4+}}}{-(p_1-p_3)^2-M^TM}NM\,.
\nonumber
\end{align}
Using the on-shell conditions and energy-momentum conservation we find, restricting for simplicity to the case of equal fermion masses $m_i=m_j=m$,
\begin{align}
\label{eq:A3-final}
\mathcal A_3^{ij}=&
\tfrac{i}{4mg}(1-p_{1+}^2)(1-p_{2+}^2)(p_{1+}p_{2+})^{-3/2}
\\
&\times\Big[
(p_{3+}-p_{4+})u^{(i}\Gamma^-\mathcal M_s(x)u^{j)}
\nonumber\\
&\qquad\qquad{}
+(p_{1+}-p_{2+})u^{[i}\Gamma^-\mathcal M_a(x)u^{j]}
\Big]\,,
\nonumber
\end{align}
where
\begin{align}
\mathcal M_s=&
NM\frac{x+2(1-m^2+M^TM)}{(1-m^2+M^TM)^2+xM^TM}NM
\nonumber\\
&{}
+M^TN^T\frac{1-m^2+M^TM}{(1-m^2+M^TM)^2+xM^TM}NM
\nonumber\\
&{}
+m^2N\frac{1-m^2+MM^T}{(1-m^2+MM^T)^2+xMM^T}N^T\,,
%
\\
\mathcal M_a=&
NM\frac{x+4(1-m^2)}{(1-m^2+M^TM)^2+xM^TM}NM
\end{align}
and we have introduced the convenient variable $x=-(p_1+p_2)^2-4=(p_{1+}-p_{2+})^2/(p_{1+}p_{2+})$, the centre-of-mass energy minus 4. The total amplitude for $z_1z_1\rightarrow\theta\theta$ scattering is then given by the sum of (\ref{eq:A3-final}) and (\ref{eq:A4-final}). Unless the mass of the fermions is also $1$ this amplitude has to vanish for factorized scattering to be possible. We can extract a relatively simple condition on the RR fluxes for this to happen by focusing on the high-energy limit $x\rightarrow\infty$ by setting $p_{1+}=1+\epsilon$, $p_{2+}=\epsilon$, $p_{3+}=1+(2-m^2)\epsilon$ and $p_{4+}=m^2\epsilon$, so that $x=\epsilon^{-1}$, and taking $\epsilon\rightarrow0$. In this limit we find that the condition becomes
\begin{equation}
\boxed{m_i^2\delta^{ij}+u^i\Gamma^-NM\frac{1}{M^TM}NMu^j=0\,.}
\label{eq:int-cond}
%
\end{equation}
Note that this condition involves only the RR fluxes, through $M$ and $N$ defined in (\ref{eq:MN}), and constant matrices. In fact we can remove $u^i$ and $u^j$ and write the LHS simply as $M^TM+NM\frac{1}{M^TM}NM$ but in that case one must remember to remove the projection onto fermions of mass 1. In deriving this condition we set the masses of the fermions to be equal but it is not hard to show that the condition takes the same form for unequal masses. In the next section we will see that this condition is in fact quite strong and can be used to rule out integrability for many backgrounds.

First let us caution that in calculating the amplitude we have ignored possible IR-divergences which can appear when there are massless fermions in the spectrum. In cases with massless fermions one therefore has to be more careful and it is possible that (\ref{eq:int-cond}) gets corrected. Luckily cases with massless fermions are very special and in fact one can often avoid dealing with massless fermions all together by picking a suitable BMN geodesic as we will see below.

\section{Examples}
\noindent $\mathbf{AdS_3\times S^3\times S^3\times S^1}$. It is instructive to see how a background which is known to be integrable manages to satisfy (\ref{eq:int-cond}). An interesting and quite non-trivial example is to take $AdS_3\times S^3\times S^3\times S^1$ but pick a non-standard (non-supersymmetric) BMN geodesic which involves an angle on both $S^3$'s \cite{Rughoonauth:2012qd}. We take the geodesic given by $x^+=\frac12(x^0+ax^5+bx^8)$ with $a^2+b^2=1$. For the type IIA solution the RR bispinor takes the form \cite{Rughoonauth:2012qd}
\begin{equation}
\mathcal S=-2\Gamma^{0129}(1-\sqrt\alpha\Gamma^{012345}-\sqrt{1-\alpha}\Gamma^{012678})\,,
\end{equation}
where the parameter $\alpha$ controls the relative size of the two $S^3$'s and we have set the AdS radius to unity $R=1$. We define rotated directions $\Gamma^{5'}=b\Gamma^5-a\Gamma^8$, $\Gamma^{8'}=a\Gamma^5+b\Gamma^8$ so that $\Gamma^\pm=\Gamma^0\pm\Gamma^{8'}$. From the definition in (\ref{eq:MN}) we then find
\begin{align}
M=&\tfrac{i}{2}\Gamma^{29}(1+a\sqrt\alpha\Gamma^{1234}+b\sqrt{1-\alpha}\Gamma^{1267})\,,\\
N=&-\tfrac12\Gamma^{345'9}(b\sqrt\alpha+a\sqrt{1-\alpha}\Gamma^{3467})\,.
\end{align}
From the fact that $M^TM=\frac14(1+a\sqrt\alpha\Gamma^{1234}+b\sqrt{1-\alpha}\Gamma^{1267})^2$ it follows that the mass spectrum consists of four different masses $m_{\pm\pm}=\frac12(1\pm a\sqrt\alpha\pm b\sqrt{1-\alpha})$ with eigenvectors $u^{\pm\pm}=\frac12(1\pm\Gamma^{1234})\frac12(1\pm\Gamma^{1267})u^{\pm\pm}$. Note that generically the masses are non-zero and also not equal to $1$, the mass of the AdS bosons. Using the fact that $a^2+b^2=1$ it is not hard to prove the nice identity $N^2u^{\pm\pm}=m_{\pm\pm}(1-m_{\pm\pm})u^{\pm\pm}$. Using this identity and the fact that $M$ and $N$ anti-commute one finds that the LHS of (\ref{eq:int-cond}) becomes
\begin{equation}
m_{++}^2
-m_{++}m_{--}u^{++}\Gamma^-M\frac{1}{M^TM}Mu^{++}
=
0\,.
\end{equation}
Similar calculations show that the remaining components of this condition are indeed also satisfied. This is consistent with the classical integrability of the string in this background \cite{Babichenko:2009dk,Sundin:2012gc}. For a proposed exact S-matrix see \cite{Borsato:2015mma}.

\vspace{.2cm}

\noindent $\mathbf{AdS_4\times S^3\times S^3}$. This example is one of the symmetric space solutions found in \cite{Wulff:2017zbl}. It is a non-supersymmetric solution of massive type IIA and the RR bispinor takes the form
\begin{equation}
\mathcal S=\sqrt2R^{-1}(1-\sqrt5\Gamma^{0123})\,.
\end{equation}
The definition (\ref{eq:MN}) implies $M=i\frac{\sqrt{10}}{4}\Gamma^{23}$ and $N=\frac{\sqrt2}{4}$. It follows that all fermions have $m^2=5/8$. The LHS of the condition (\ref{eq:int-cond}) becomes $3/4$ which does not vanish and therefore integrability is ruled out for this background. Note however that the bosonic string is integrable since we are dealing with a symmetric space and there is no NSNS flux.

\vspace{.2cm}

\noindent$\mathbf{AdS_3\times S^3\times S^2\times H^2}$. This example is another of the symmetric space solutions of \cite{Wulff:2017zbl}, see also \cite{Figueroa-OFarrill:2012whx}. It is a non-supersymmetric type IIB solution with RR bispinor
\begin{equation}
\mathcal S=\tfrac{i}{2}\sigma^2\big(f_3(\gamma^{01289}-\gamma^{34567})+f_4(\gamma^{01267}-\gamma^{34589})\big)\,,
\end{equation}
where the AdS radius is given by $R^{-2}=(f_3^2+f_4^2)/8$. To avoid massless fermions it is convenient to take the BMN geodesic given by $x^+=\frac12(x^0+x^{9'})$, where we've made a rotation in the (79)-plane to $x^{7'}=\frac{1}{\sqrt2}(x^7-x^9)$, $x^{9'}=\frac{1}{\sqrt2}(x^7+x^9)$. From the definition in (\ref{eq:MN-IIB}) we find
\begin{align}
M=&\tfrac{iR}{8\sqrt2}\sigma^2\gamma^{27'}(f_3\gamma^8-f_4\gamma^6)(1-\gamma^{1234567'8})\,,
\\
N=&\tfrac{R}{8\sqrt2}\sigma^2\gamma^{12}(-f_3\gamma^8-f_4\gamma^6)(1+\gamma^{1234567'8})\,.
\end{align}
From the first expression we find $M^TM=\frac{1}{8}(1-\gamma^{1234567'8})$ and noting that $\gamma^{1234567'8}u^i=\gamma^{0123456789}u^i=-u^i$ we find that all fermions have $m=\frac12$. The LHS of (\ref{eq:int-cond}) becomes
\begin{equation}
\frac14+\frac{R^4}{256}u\gamma^-(f_3^2-f_4^2+2f_3f_4\gamma^{68})^2u
=
\frac12-\frac{2f_3^2f_4^2}{(f_3^2+f_4^2)^2}\,.
\end{equation}
For this to vanish we must have $f_4=\pm f_3$ but in this case the background degenerates to $AdS_3\times S^3\times T^4$. Therefore integrability is ruled out for this background. Since the RR flux for the backgrounds $AdS_3\times S^5\times H^2$ and $AdS_3\times SLAG_3\times H^2$ is of the same form, but with $f_4=0$, integrability is ruled out also for these backgrounds.

\section{Conclusions}
We have used the fact that factorization of worldsheet scattering requires many two-to-two amplitudes to vanish, namely those for which the set of initial an final masses and momenta differ, to constrain the RR fluxes of integrable AdS supergravity backgrounds. In particular we have found the constraint (\ref{eq:int-cond}) with $M,N$ determined from the RR fluxes by (\ref{eq:MN}) and (\ref{eq:SIIA}) or the corresponding type IIB expressions. We have also seen how this condition can be used to rule out integrability for some of the symmetric space backgrounds of \cite{Wulff:2017zbl}. In a forthcoming publication we will extend this to rule out integrability for the remaining non-supersymmetric backgrounds of \cite{Wulff:2017zbl}.

We hope to also apply this condition, or a suitable modification, to more complicated backgrounds which are not of symmetric space form. Fortunately there is a vast literature on (supersymmetric) AdS backgrounds to exploit. In this way we expect to be able to constrain severely the space of integrable AdS/CFT-pairs.



\bibliographystyle{h-physrev}
\bibliography{biblio}

\begin{thebibliography}{10}

\bibitem{Maldacena:1997re}
J.~M. Maldacena,
\newblock Int.J.Theor.Phys. {\bf 38}, 1113 (1999), hep-th/9711200.

\bibitem{Bena:2003wd}
I.~Bena, J.~Polchinski, and R.~Roiban,
\newblock Phys.Rev. {\bf D69}, 046002 (2004), hep-th/0305116.

\bibitem{Minahan:2002ve}
J.~Minahan and K.~Zarembo,
\newblock JHEP {\bf 0303}, 013 (2003), hep-th/0212208.

\bibitem{Beisert:2010jr}
N.~Beisert {\em et~al.},
\newblock Lett. Math. Phys. {\bf 99}, 3 (2012), 1012.3982.

\bibitem{Bombardelli:2016rwb}
D.~Bombardelli {\em et~al.},
\newblock J. Phys. {\bf A49}, 320301 (2016), 1606.02945.

\bibitem{Bombardelli:2009ns}
D.~Bombardelli, D.~Fioravanti, and R.~Tateo,
\newblock J.Phys. {\bf A42}, 375401 (2009), 0902.3930.

\bibitem{Gromov:2009bc}
N.~Gromov, V.~Kazakov, A.~Kozak, and P.~Vieira,
\newblock Lett.Math.Phys. {\bf 91}, 265 (2010), 0902.4458.

\bibitem{Arutyunov:2009ur}
G.~Arutyunov and S.~Frolov,
\newblock JHEP {\bf 0905}, 068 (2009), 0903.0141.

\bibitem{Gromov:2013pga}
N.~Gromov, V.~Kazakov, S.~Leurent, and D.~Volin,
\newblock Phys.Rev.Lett. {\bf 112}, 011602 (2014), 1305.1939.

\bibitem{Aharony:2008ug}
O.~Aharony, O.~Bergman, D.~L. Jafferis, and J.~Maldacena,
\newblock JHEP {\bf 0810}, 091 (2008), 0806.1218.

\bibitem{Arutyunov:2008if}
G.~Arutyunov and S.~Frolov,
\newblock JHEP {\bf 09}, 129 (2008), 0806.4940.

\bibitem{Stefanski:2008ik}
B.~Stefanski, jr,
\newblock Nucl. Phys. {\bf B808}, 80 (2009), 0806.4948.

\bibitem{Sorokin:2010wn}
D.~Sorokin and L.~Wulff,
\newblock JHEP {\bf 11}, 143 (2010), 1009.3498.

\bibitem{Babichenko:2009dk}
A.~Babichenko, J.~Stefanski, B., and K.~Zarembo,
\newblock JHEP {\bf 1003}, 058 (2010), 0912.1723.

\bibitem{Sorokin:2011rr}
D.~Sorokin, A.~Tseytlin, L.~Wulff, and K.~Zarembo,
\newblock J. Phys. {\bf A44}, 275401 (2011), 1104.1793.

\bibitem{Wulff:2014kja}
L.~Wulff,
\newblock JHEP {\bf 05}, 115 (2014), 1402.3122.

\bibitem{Basu:2011di}
P.~Basu and L.~A. Pando~Zayas,
\newblock Phys. Lett. {\bf B700}, 243 (2011), 1103.4107.

\bibitem{Stepanchuk:2012xi}
A.~Stepanchuk and A.~A. Tseytlin,
\newblock J. Phys. {\bf A46}, 125401 (2013), 1211.3727.

\bibitem{Chervonyi:2013eja}
Y.~Chervonyi and O.~Lunin,
\newblock JHEP {\bf 02}, 061 (2014), 1311.1521.

\bibitem{Giataganas:2013dha}
D.~Giataganas, L.~A. Pando~Zayas, and K.~Zoubos,
\newblock JHEP {\bf 01}, 129 (2014), 1311.3241.

\bibitem{Wulff:2017hzy}
L.~Wulff,
\newblock J. Phys. {\bf A50}, 23LT01 (2017), 1702.08788.

\bibitem{Zamolodchikov:1978xm}
A.~B. Zamolodchikov and A.~B. Zamolodchikov,
\newblock Annals Phys. {\bf 120}, 253 (1979).

\bibitem{Cvetic:1999zs}
M.~Cvetic, H.~Lu, C.~Pope, and K.~Stelle,
\newblock Nucl.Phys. {\bf B573}, 149 (2000), hep-th/9907202.

\bibitem{Wulff:2013kga}
L.~Wulff,
\newblock JHEP {\bf 1307}, 123 (2013), 1304.6422.

\bibitem{Berenstein:2002jq}
D.~E. Berenstein, J.~M. Maldacena, and H.~S. Nastase,
\newblock JHEP {\bf 0204}, 013 (2002), hep-th/0202021.

\bibitem{Rughoonauth:2012qd}
N.~Rughoonauth, P.~Sundin, and L.~Wulff,
\newblock JHEP {\bf 07}, 159 (2012), 1204.4742.

\bibitem{Sundin:2012gc}
P.~Sundin and L.~Wulff,
\newblock JHEP {\bf 10}, 109 (2012), 1207.5531.

\bibitem{Borsato:2015mma}
R.~Borsato, O.~O. Sax, A.~Sfondrini, and B.~Stefa\'nski,
\newblock (2015), 1506.00218.

\bibitem{Wulff:2017zbl}
L.~Wulff,
\newblock (2017), 1706.02118.

\bibitem{Figueroa-OFarrill:2012whx}
J.~Figueroa-O'Farrill and N.~Hustler,
\newblock Class. Quant. Grav. {\bf 30}, 045008 (2013), 1209.4884.

\bibitem{Fradkin:1984pq}
E.~S. Fradkin and A.~A. Tseytlin,
\newblock Phys. Lett. {\bf 158B}, 316 (1985).

\bibitem{Fradkin:1985ys}
E.~S. Fradkin and A.~A. Tseytlin,
\newblock Nucl. Phys. {\bf B261}, 1 (1985),
\newblock [Erratum: Nucl. Phys.B269,745(1986)].

\bibitem{Wulff:2016tju}
L.~Wulff and A.~A. Tseytlin,
\newblock JHEP {\bf 06}, 174 (2016), 1605.04884.

\end{thebibliography}

\end{document}